\documentclass[sigconf]{acmart}
\AtBeginDocument{%
  }

\usepackage{algorithmic}
\usepackage{algorithm}
\usepackage{amsthm}
\usepackage{appendix}
\usepackage{amsfonts}
\usepackage{multirow}
\usepackage{subcaption}
\usepackage{url}
\usepackage{nicefrac}

\graphicspath{ {./images/} }
\newtheorem{assumption}{Assumption}

\copyrightyear{2026}
\acmYear{2026}
\setcopyright{cc}
\setcctype{by}
\acmDOI{10.1145/3847352.3848120}
\acmConference[AISec '26]{19th Workshop on Artificial Intelligence and Security}{November 15--19, 2026}{The Hague, Netherlands}
\acmBooktitle{19th Workshop on Artificial Intelligence and Security (AISec '26), November 15--19, 2026, The Hague, Netherlands}
\acmISBN{979-8-4007-3031-3/2026/11}

\begin{document}

\title{Calibrating One-Round Membership Inference with Neighbors}

\author{Francesco Rita}
\affiliation{
  \institution{ETH Zurich}
  \city{Zurich}
  \country{Switzerland}
}

\author{Jie Zhang}
\affiliation{
  \institution{ETH Zurich}
  \city{Zurich}
  \country{Switzerland}
}

\author{Florian Tramèr}
\affiliation{
  \institution{ETH Zurich}
  \city{Zurich}
  \country{Switzerland}
}

\begin{abstract}
The state-of-the-art Membership Inference (MI) methods calibrate their signal separately for each example using reference models, auxiliary models trained to exclude the target. This paradigm scales poorly to modern large models, however, whose training is too expensive to replicate. This has motivated one-round settings, where only a single trained model is available; but without reference models the per-example calibration that drives the strongest attacks can no longer be estimated, leaving the membership signal weak. We ask whether neighbors of the target point can recover this calibration without training any additional model. Our key observation is that reference models serve only to reveal how an example behaves under models not trained on it, and that querying the target model on nearby samples yields the same information. We propose two complementary ways to obtain such neighbors, and show that querying them against an early training checkpoint further sharpens the signal. 
We evaluate across three image classification datasets and three training setups, showing that neighbors yield strong membership signals and competitive attack performance at no additional training cost.
\end{abstract}



\begin{teaserfigure}
    \includegraphics[width=\linewidth]{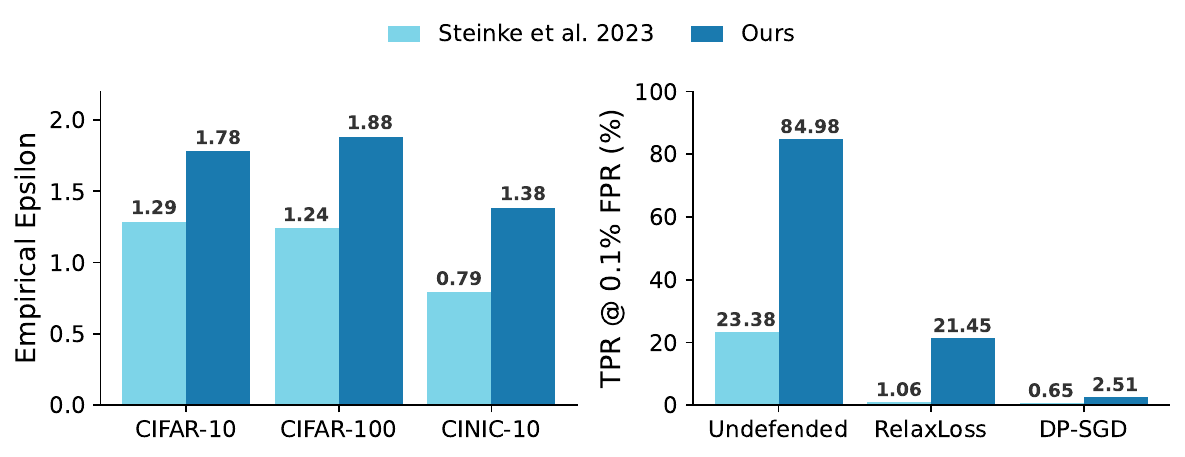}
    \caption{Comparison between \citet{steinke2023privacy} and our best method. On the left we report the empirical $\epsilon$ in the DP-SGD setting across CIFAR-10, CIFAR-100 and CINIC-10. On the right we report the TPR at 0.1\% FPR on CIFAR-10 across three privacy settings: undefended models, RelaxLoss and DP-SGD.}
    \Description{Two bar plots illustrating the improvement in privacy leakage achieved by our best-performing method compared with \citet{steinke2023privacy}. The left panel reports the empirical $\epsilon$ in the DP-SGD setting across CIFAR-10, CIFAR-100, and CINIC-10. The right panel shows the true positive rate (TPR) at a 0.1\% false positive rate (FPR) on CIFAR-10 under three privacy settings: undefended models, RelaxLoss, and DP-SGD. Across all datasets and privacy settings, our method consistently achieves higher privacy leakage than the baseline.}
    \label{fig:steinke_vs_synth_ec}
\end{teaserfigure}

\maketitle

\section{Introduction}
Membership inference attacks~\cite{shokri2017membership} have become a
standard tool for auditing the privacy of machine learning models. Given a trained
model and a target sample, an MIA decides whether that sample was part of the
model's training set. Because models tend to behave differently on data they have
memorized than on unseen data, the success of such an attack offers an empirical,
sample-level measure of how much a model leaks about its training data.

The strongest MIAs achieve this by \emph{calibrating} the target's behavior on a
per-example basis. Rather than asking whether the loss on a sample is low in
absolute terms, they ask whether it is surprisingly low \emph{for that particular
sample}, accounting for the fact that some examples are intrinsically easier than
others~\cite{carlini2022lira}. This calibration is precisely what separates
state-of-the-art attacks from naive loss thresholding, and it is obtained through
\emph{reference} (or \emph{shadow}) models: auxiliary models that reveal how the
target model would score a sample when that sample is excluded from
training~\cite{shokri2017membership, carlini2022lira, du2025cascadingproxymembershipinference, bai2025efficientinferenceattacksshadow}.

This paradigm, however, has grown increasingly impractical with the advent of
highly scalable architectures such as Transformers~\cite{vaswani2023attentionneed}
and the large foundation models they
enabled~\cite{touvron2023llamaopenefficientfoundation, deepseekai2025deepseekv3technicalreport}.
Training even a single such model can be prohibitively expensive; training the tens
or hundreds of reference models that calibration requires is often out of reach.
This has motivated growing interest in the \emph{one-round} setting, where the
auditor has access to a single trained model and cannot retrain new models. The one-round
regime is especially central to the auditing of differentially private (DP)
training, where one seeks an empirical lower bound on the privacy parameter
$\epsilon$ from a single training run, precisely to avoid the cost of repeated
retraining~\cite{steinke2023privacy, mahloujifar2024auditingfdifferentialprivacyrun, xiang2025tightprivacyauditrun}.

\begin{figure*}[t]
    \centering
\includegraphics[width=1\textwidth]{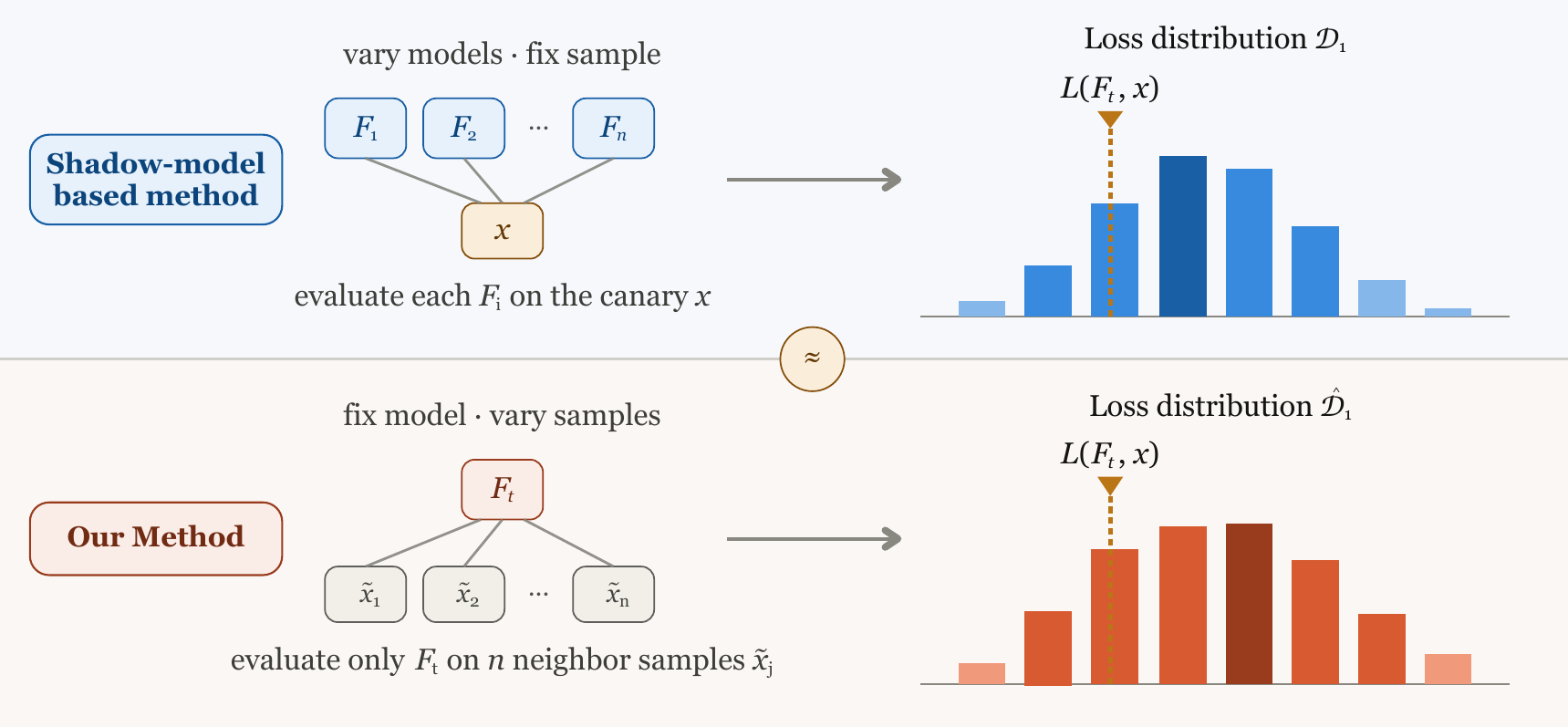}
    \caption{Comparison between shadow-model based methods and our method. Instead of characterizing the loss distribution by querying the canary against different shadow models, our method queries different samples neighboring the canary against the target model alone.}
    \Description{Illustration of the key difference between the shadow-model-based approach and our method. The traditional approach fixes a target sample and evaluates it across multiple shadow models to construct a loss distribution. Our method instead fixes the target model and evaluates it on multiple neighboring samples of the target, yielding an approximation of the same loss distribution without the need to train or query multiple models.}
    \label{fig:method_comparison}
\end{figure*}

Removing reference models, however, removes the very ingredient that makes modern
attacks strong. Without them, the per-example calibration can no longer be
estimated, and existing one-round methods fall back on a global loss threshold or
simple heuristics~\cite{steinke2023privacy}. The resulting signal is weak, particularly at the low
false-positive rates that matter most for membership
inference~\cite{carlini2022lira}, which in turn loosens DP audits. Recovering
per-example calibration \emph{without} reference models is therefore the common
bottleneck for both one-round attacks and one-run DP auditing.

In this work, we show that this calibration can be recovered using nothing beyond the
target model itself. Our starting point is the observation that reference models
serve a single purpose: to reveal how the target sample would score under models
that did not train on it. We obtain the same information differently, by querying
the target model on samples that \emph{neighbor} the target point. Intuitively, the
model's behavior on a non-member sample close to the target approximates the
behavior a reference model would exhibit on the target itself.
Figure~\ref{fig:method_comparison} contrasts the two views: where reference-based attacks query the target sample against many models, we query many neighbors against a single model. Building on this idea, we characterize what makes a neighbor useful
and propose two complementary ways to construct neighbors, by retrieving them from
held-out data (\emph{Similarity Neighbors}) and by synthesizing them with a
diffusion model (\emph{Synthetic Neighbors}); in the context of internal audits, we further show that querying neighbors
against an \emph{early training checkpoint} sharpens the signal further. Across three
datasets and three privacy regimes, our attacks substantially outperform the
one-round baseline of \citet{steinke2023privacy} and close much of the gap to
reference-based attacks, without training a single additional model, as previewed in
Figure~\ref{fig:steinke_vs_synth_ec}.

\section{Background and Preliminary}
\label{sec:background}

\subsection{Membership Inference: Setup and Threat Model}
\label{subsec:mia-setup}

Let $F$ denote a model trained on a dataset $D$ drawn from a data
distribution $\mathbb{D}$. \emph{Membership inference} (MI) asks, for a
given example $x$, whether $x \in D$. Following the standard
formulation~\cite{shokri2017membership, yeom2018privacy}, an MI
adversary $\mathcal{A}$ outputs a score reflecting its belief that $x$
was a training member; thresholding this score yields a membership
decision, and the score's quality is measured by the resulting
true-positive/false-positive trade-off.  Within this framework, we refer
to \emph{canaries} as the subset of training examples that are most
vulnerable to membership inference.

\paragraph{Threat model.} We study the most constrained and most
realistic regime, which we refer to as \emph{one-round} membership
inference. The adversary is given a single target model $F_t$ and an
example $x$, and may query $F_t$ to obtain a per-example signal such as
the loss or confidence on $x$. Crucially, the adversary \emph{cannot}
train additional models. This rules out the reference-model machinery on
which the strongest attacks rely, and it is the setting faced by an
auditor or adversary confronting a single deployed model.

\subsection{From Loss Thresholding to Reference-Based Attacks}
\label{subsec:loss-to-shadow}

The simplest attack thresholds the per-example loss: if $L(F_t, x)$ is
small, predict ``member''~\cite{yeom2018privacy}. This is weak because
examples differ in intrinsic difficulty. Some examples have low loss
whether or not they were trained on, while others have high loss
regardless, so a single global threshold conflates easy non-members
with true members. The membership signal comes not from the magnitude
of the loss but from \emph{calibrating} that loss to the specific
example.

Reference-model (or \emph{shadow-model}) attacks supply this
calibration~\cite{shokri2017membership, carlini2022lira}. By training
auxiliary models on data drawn from $\mathbb{D}$, the adversary
estimates how a model behaves on $x$ when $x$ is, or is not, in its
training set, and calibrates the target's behavior against this
reference. The Likelihood Ratio Attack
(LiRA)~\cite{carlini2022lira} makes this precise: it casts membership
as a per-example hypothesis test, comparing the target's score on $x$
against the distribution of scores that models \emph{not} trained on
$x$ assign to $x$.

\subsection{Offline LiRA, Formally}
\label{subsec:offline_lira}

Let $s(F, x)$ be a membership score for example $x$ under model $F$. A
standard choice is the logit-scaled confidence~\cite{carlini2022lira}
\begin{equation}
  s(F, x) \;=\; \log \frac{p}{1 - p},
  \qquad p = F(x)_y,
  \label{eq:score}
\end{equation}
where $F(x)_y$ is the predicted probability on the true label $y$;
larger $s$ indicates stronger memorization of $x$ by $F$. The
logit scaling is used because $s$ is approximately Gaussian, whereas the
raw confidence is not.

\emph{Offline} LiRA assumes access only to \textsc{out} models---models
not trained on $x$---and performs a one-sided test.\footnote{This is the
appropriate baseline for our setting: the \emph{online} variant
additionally trains \textsc{in} models that include $x$, which requires
retraining with the target example and is therefore unavailable in the
one-round regime. For brevity, we henceforth write \emph{LiRA} to denote
its offline variant unless stated otherwise.} Concretely, it trains a set
of \textsc{out} shadow models $\{F_1, \dots, F_n\}$, none of which is
trained on $x$, and approximates the \textsc{out} score distribution as a
Gaussian
\begin{equation}
  \mathcal{D}_1 \;=\; \mathcal{N}\!\left(\mu_{\mathrm{out}},\,
  \sigma_{\mathrm{out}}^2\right),
  \label{eq:out-dist}
\end{equation}
with
\begin{equation}
  \mu_{\mathrm{out}} = \frac{1}{n} \sum_{i=1}^{n} s(F_i, x),
  \qquad
  \sigma_{\mathrm{out}}^2 = \frac{1}{n} \sum_{i=1}^{n}
  \big( s(F_i, x) - \mu_{\mathrm{out}} \big)^2 .
  \label{eq:out-params}
\end{equation}
Membership is then assessed by how surprising the target's score is
under $\mathcal{D}_1$:
\begin{equation}
  \Lambda(x) \;=\; 1 - \Phi\!\left(
  \frac{s(F_t, x) - \mu_{\mathrm{out}}}{\sigma_{\mathrm{out}}}
  \right),
  \label{eq:offline_lira}
\end{equation}
where $\Phi$ is the standard normal CDF. A small $\Lambda(x)$---the
target's score lying in the upper tail of $\mathcal{D}_1$---indicates
that $x$ is a member.

The entire procedure reduces to one object: the per-example
\textsc{out} distribution $\mathcal{D}_1$, estimated from a set of
\textsc{out} scores $\{s(F_i, x)\}_{i=1}^{n}$. Its strength comes
entirely from this per-example calibration, and its cost comes entirely
from obtaining the $n$ \textsc{out} models needed to form it.

\subsection{One-Round MI and One-Run DP Auditing}
\label{subsec:one-round-auditing}

The reference models underlying Eq.~\eqref{eq:out-params} are precisely
what the one-round setting denies. Without the ability to train
\textsc{out} models, $\mathcal{D}_1$ cannot be estimated per example,
and existing one-round methods fall back on a global loss threshold or
simple per-sample heuristics---discarding the calibration that drives
the signal. The result is a faint membership signal, especially in the
low-false-positive-rate regime that matters most for
MI~\cite{carlini2022lira}.

The same constraint motivates the auditing of differentially private
(DP) training~\cite{dwork2006calibrating}. DP auditing derives an
empirical lower bound on the privacy parameter $\varepsilon$ from the
distinguishability of members and non-members, and has moved decisively
toward the \emph{single-run} regime~\cite{steinke2023privacy}: rather
than retraining a model many times, one inserts many canary examples
into a single training run and infers $\varepsilon$ from their
membership scores, precisely to avoid the prohibitive cost of repeated
training. The tightness of such an audit---how close the empirical
$\varepsilon$ comes to the true privacy loss---is governed by the
separation between member and non-member scores. When each canary is
scored by its global loss, this separation is poor, and the audit is
correspondingly loose, forcing either an underestimate of $\varepsilon$
or an impractically large number of canaries.

Recovering per-example calibration \emph{without} \textsc{out} models is
therefore the common bottleneck for both one-round attacks and one-run
auditing. In Section~\ref{sec:method}, we address it directly: we
estimate the \textsc{out} distribution $\mathcal{D}_1$ using the target
model's scores on \emph{neighbors} of $x$, requiring nothing beyond the
target model itself.
\section{Calibrating One-Round MI with Neighbors}
\label{sec:method}

LiRA owes its strength to a single object: the per-example
\textsc{out} distribution $\mathcal{D}_1$ of Eq.~\eqref{eq:out-dist},
whose mean and variance are estimated from the scores that \textsc{out}
shadow models assign to the example under test---the \emph{canary}, in
the auditing setting. As noted in
Section~\ref{subsec:offline_lira}, the shadow models serve no purpose
beyond producing this set of scores. We exploit this observation
directly: rather than training shadow models, we form the same set of
scores by querying the target model $F_t$ on samples that
\emph{neighbor} the canary (see
Figure~\ref{fig:method_comparison}). The intuition, which we make precise
below, is that the target's behavior on nearby non-member samples
proxies the canary's behavior across \textsc{out} models.

Formally, our attack is a single substitution in the LiRA
estimator. Given a set of $n$ neighbors $\{\tilde{x}_1,\dots,\tilde{x}_n\}$
of the canary $x$, we replace the \textsc{out}-model scores
$\{s(F_i, x)\}$ with the target's scores on the neighbors and estimate
\begin{equation}
  \hat{\mu}_{\mathrm{out}} = \frac{1}{n}\sum_{j=1}^{n} s(F_t, \tilde{x}_j),
  \qquad
  \hat{\sigma}^2_{\mathrm{out}} = \frac{1}{n}\sum_{j=1}^{n}
  \big( s(F_t, \tilde{x}_j) - \hat{\mu}_{\mathrm{out}} \big)^2 ,
  \label{eq:neighbor-estimator}
\end{equation}
yielding $\hat{\mathcal{D}}_1 = \mathcal{N}(\hat{\mu}_{\mathrm{out}},
\hat{\sigma}^2_{\mathrm{out}})$. We then score the canary by the tail
probability of $s(F_t, x)$ under $\hat{\mathcal{D}}_1$, exactly as in LiRA (Eq.~\eqref{eq:offline_lira}). The full procedure is given
in Algorithm~\ref{alg:our_method}; it requires no model training and
queries only $F_t$, making it applicable in the one-round setting.

\begin{figure}[h]
    \centering
    \includegraphics[width=\linewidth]{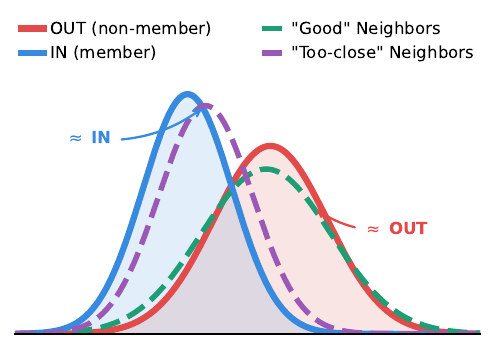}
    \caption{Comparison of shadow model-based and neighbor-driven confidence distributions for a member canary. The \textsc{in} (blue) and \textsc{out} (red) distributions are estimated using shadow models trained with and without the canary, respectively. Neighbor distributions are obtained by querying Stable Diffusion img2img generations against the target model. Neighbors that are too similar (strength 0.45) resemble the member (\textsc{in}) distribution, whereas stronger perturbations (strength 0.75) better approximate the non-member (\textsc{out}) distribution.}
    \Description{Comparison of confidence distributions obtained from shadow models and synthetic neighbors for a member sample. Neighbors that are too similar to the target sample produce confidence values that closely follow the shadow-model \textsc{in} distribution, whereas sufficiently perturbed neighbors better approximate the \textsc{out} distribution.}
    \label{fig:lira_distributions}
\end{figure}

\subsection{What Makes a ``Good'' Neighbor}
\label{subsec:good_neighbors}

The validity of the substitution in Eq.~\eqref{eq:neighbor-estimator}
rests on one assumption, which we state explicitly.

\begin{assumption}[Local smoothness]
\label{ass:smoothness}
The target model's membership score $s(F_t, \cdot)$ varies smoothly over
the data manifold in a neighborhood of the canary $x$. Consequently, for
a non-member $\tilde{x}$ sufficiently close to $x$, the score
$s(F_t, \tilde{x})$ approximates the score an \textsc{out} model assigns
to $x$, i.e.\ a sample from $\mathcal{D}_1$.
\end{assumption}

Assumption~\ref{ass:smoothness} exposes a distance trade-off intrinsic to
our estimator. To faithfully approximate $\mathcal{D}_1$, each neighbor
must be close enough to the canary to be representative of its behavior.
Yet being too close is harmful: if the canary was seen during training,
the target memorizes it, and a neighbor that is too similar has its score
pulled toward the member regime, contaminating
$\hat{\mathcal{D}}_1$ and shrinking the very membership signal we seek.
For member canaries the trade-off depends further on intrinsic learning
difficulty and the model's memorization capacity: a hard-to-memorize
sample exhibits similar scores regardless of membership, tolerating
closer neighbors, whereas an easy-to-memorize sample has a wide
membership-dependent score gap, requiring neighbors to lie further away.

Because membership labels are unavailable at inference time, this
trade-off must be respected whether or not the canary is a member. Even
if one could condition on membership, the optimal distance would still
depend on canary- and model-specific characteristics that vary
considerably across samples and settings. These factors make it
difficult to formulate a single distance threshold whose tuning reliably
optimizes the trade-off across all canaries. We therefore do not attempt
to pin down one optimal distance. Instead, we keep neighbors within a
controlled similarity \emph{band}---close to the canary, yet not too
close---and operationalize this band through two complementary
construction mechanisms: the intrinsic train--test gap when retrieving
neighbors from the test set (Section~\ref{subsec:similarity_neighbors}),
and the denoising strength when synthesizing them
(Section~\ref{subsec:synthetic_neighbors}). Both mechanisms use a notion
of distance to \emph{locate} candidates near the canary, but neither
relies on a single tuned distance threshold as its decision criterion.

\begin{algorithm}[t!]
\caption{Neighbor approximation of LiRA. In the Early Checkpoint
variant, the neighbor scores in line~\ref{instr:neighbor_losses} are
computed against an early checkpoint of the target model.}
\label{alg:our_method}
\begin{algorithmic}[1]
    \REQUIRE target model $F_t$, canary $x$, number of neighbors $n$
    \STATE $\{\tilde{x}_1, \dots, \tilde{x}_n\} \leftarrow \mathtt{neighbors}(x)$ \hfill
      \COMMENT{\textit{search / generate neighbors}}
    \STATE $\mathrm{S_{out}} \leftarrow \{ s(F_t, \tilde{x}_1), \dots, s(F_t, \tilde{x}_n) \}$
      \label{instr:neighbor_losses}
    \STATE $\hat{\mu}_{\mathrm{out}} \leftarrow \mathtt{mean}(\mathrm{S_{out}})$
    \STATE $\hat{\sigma}^2_{\mathrm{out}} \leftarrow \mathtt{var}(\mathrm{S_{out}})$
    \RETURN $1 - \mathtt{GaussianTail}\big(s(F_t, x);\,
      \hat{\mu}_{\mathrm{out}}, \hat{\sigma}^2_{\mathrm{out}}\big)$
\end{algorithmic}
\end{algorithm}

\subsection{Similarity Neighbors}
\label{subsec:similarity_neighbors}

The test set is a natural source of neighbors: by construction it
consists entirely of non-member data, shares the training
distribution, and is held at the intrinsic train--test distance from any
training sample---precisely the upper edge of the band described in
Section~\ref{subsec:good_neighbors}. Exploiting this, we select each
canary's neighbors by retrieving the most similar test samples.

Concretely, given a canary we first discard all test samples not in its
class. We then embed the canary and each remaining test sample with
OpenAI's CLIP ViT-B/32 model
\cite{radford2021learningtransferablevisualmodels} and compute their
cosine similarity. To obtain a more robust score, we repeat this over $18$
augmentations of the input pair (horizontal shifts, vertical shifts, and
vertical flips) and average the resulting scores. Simpler variants that
omit the class filter or the augmentations are possible; we adopt this
formulation as it yields the best empirical performance. As shown in
Section~\ref{sec:results}, it already outperforms
\citet{steinke2023privacy} in most scenarios.

\subsection{Synthetic Neighbors}
\label{subsec:synthetic_neighbors}

Our second construction shifts from existing data to synthetic neighbors.
This offers greater flexibility, as the generated samples can be
controlled through the choice of generative model, its hyperparameters,
and optional text prompting---and it applies even when the test set
contains no suitable neighbors. We generate neighbors with the
image-to-image pipeline of Stable
Diffusion\footnote{We use the
\href{https://huggingface.co/stable-diffusion-v1-5/stable-diffusion-v1-5}{\texttt{stable-diffusion-v1-5}}
implementation, as the original \texttt{RunwayML} model is deprecated.}
\cite{rombach2022highresolutionimagesynthesislatent}, which produces
variations of an input image by partially noising and then denoising it.

The pipeline is governed by three core hyperparameters:
\begin{itemize}
    \item \textbf{Number of steps.} The total number of denoising
      iterations.
    \item \textbf{Strength.} A value in $[0,1]$ controlling the diffusion
      step from which denoising begins, effectively determining how much
      the output may deviate from the input.
    \item \textbf{Guidance scale.} The influence of the text prompt on
      generation.
\end{itemize}
Strength is the mechanism by which we control the similarity band of
Section~\ref{subsec:good_neighbors}: to enforce sufficient distance
between canary and neighbors, we consider only medium-to-high strength
values, granting higher variance from the original image. To assess the
potential of this strategy, we report the highest score across
\textit{number of steps} $\in \{30, 35, 40\}$, \textit{strength}
$\in \{0.55, 0.65, 0.75, 0.85\}$, and \textit{guidance scale}
$\in \{1, 3, 5\}$. For guidance scale $1$ we use an empty prompt;
for larger values we use a short class-representative prompt (e.g.,
``A dog'', ``An airplane''). As shown in Section~\ref{sec:results}, this
approach consistently outperforms both the baseline
\citet{steinke2023privacy} and our first proposal.

\begin{figure}[h]
    \includegraphics[width=\linewidth]{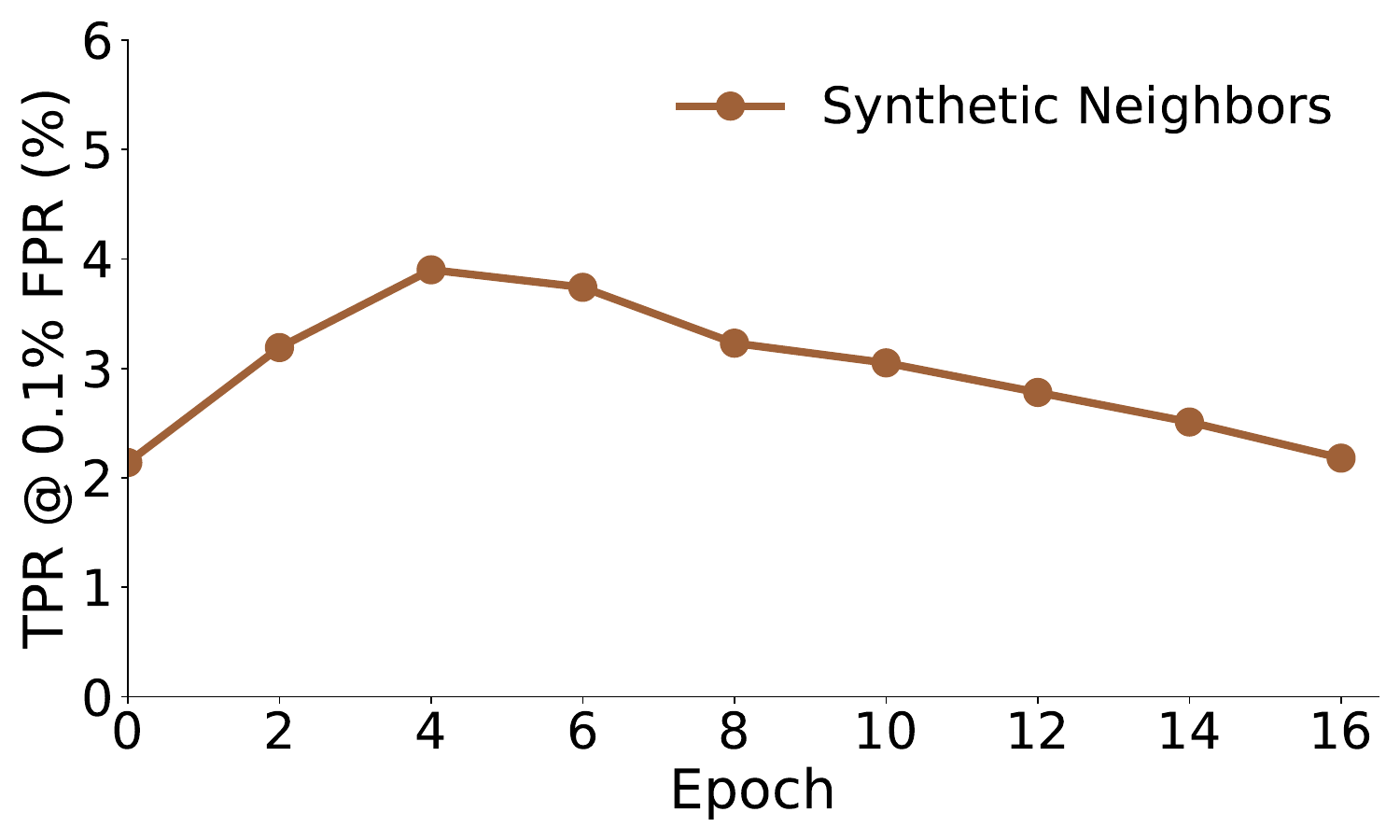}
    \caption{TPR at 0.1\% FPR of our attack when querying synthetic neighbors against early training checkpoints, from a real example.}
    \Description{Attack performance measured across successive training checkpoints for a representative sample. Rather than increasing monotonically, the TPR at 0.1\% FPR peaks during the early stages of training and subsequently declines.}
    \label{fig:synth_ec_train}
\end{figure}

\subsection{Early Checkpoints}
\label{subsec:early_checkpoints}

To ease the distance trade-off of
Section~\ref{subsec:good_neighbors}, we propose evaluating the neighbors'
scores against an \emph{early checkpoint} of the target model rather than
the fully trained one. We call this the Early Checkpoint (E.C.) variant.
Unlike our base attacks, this variant relaxes the black-box assumption, 
as it requires access to intermediate training snapshots rather than the 
fully trained model alone; it is therefore best suited to internal auditing
settings.

The intuition is that before memorization sets in, the model has acquired
generalization but has not yet overfit specific training examples. At
this stage the score landscape around a member canary resembles that of
non-member samples, so even closer neighbors serve as good non-member
representatives. Figure~\ref{fig:synth_ec_train} corroborates this 
intuition: the attack TPR peaks in the early epochs and progressively 
declines as training continues.

Since we cannot know a priori when memorization begins,
we periodically snapshot the model during training and select the
checkpoint yielding the highest privacy leakage. Crucially, the early
checkpoint is used \emph{only} to compute the neighbor scores that
characterize $\hat{\mathcal{D}}_1$; the canary score is still evaluated
against the fully trained model. In Algorithm~\ref{alg:our_method} this
amounts to replacing $F_t$ with an early checkpoint exclusively in
line~\ref{instr:neighbor_losses}.

We note that this variant relaxes the black-box assumption: it requires
access to intermediate training states. This is essentially free in the
one-run \emph{auditing} setting, where the auditor controls training and
can save checkpoints at no cost, but it is a stronger assumption for an
external \emph{attacker}, for whom it applies only when such checkpoints
are available. In Section~\ref{sec:results} we show that, where
applicable, E.C.\ consistently improves performance across all our
methods.
\section{Experiments}
Following \citet{Aerni_2024}, we adopt their model architectures and training pipelines, in accordance with their recommendation to evaluate privacy defenses on high-utility models. 
We evaluate each attack across three datasets and three privacy settings, resulting in nine experimental scenarios. 
In every scenario, the objective is to infer the membership of 500 canaries across 64 independently trained models.

Canaries are randomly selected from the training set and assigned independently to half of the models. 
Each model is trained on all non-canary samples together with its assigned canaries. 
This assignment ensures that every canary appears equally often as a member and a non-member sample across the ensemble, enabling a balanced evaluation of membership inference performance. 
To increase memorization and maximize privacy leakage, the labels of all canaries are flipped prior to training.

\begin{figure*}[h!t!]
    \centering
\includegraphics[width=0.85\textwidth]{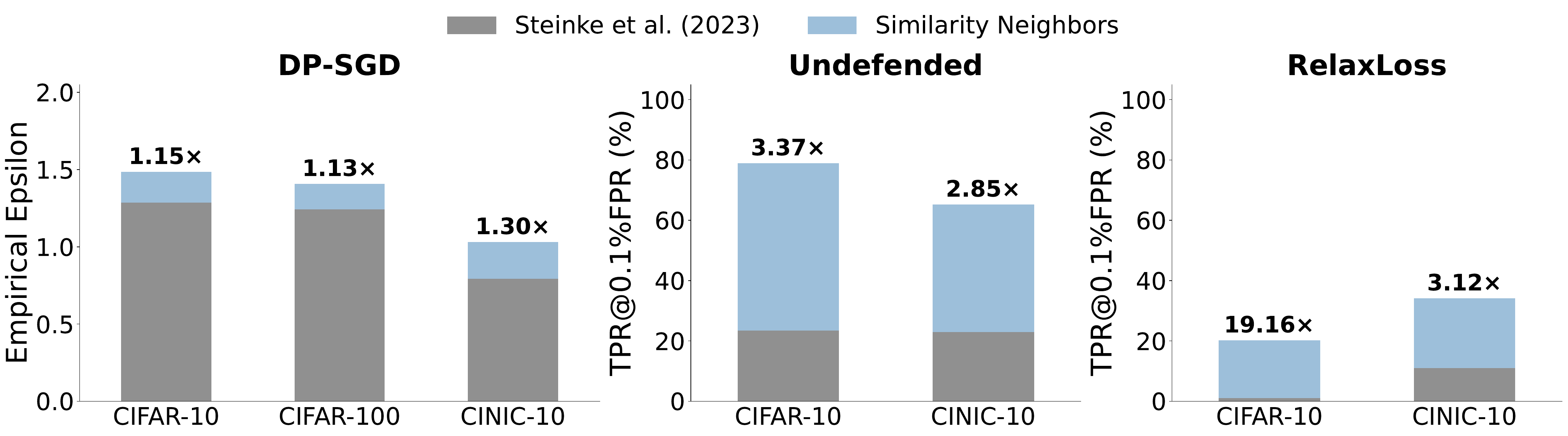}
    \caption{Settings where our Similarity Neighbors approach significantly outperforms \citet{steinke2023privacy}.}
    \Description{Three grouped bar charts comparing privacy leakage estimates from \citet{steinke2023privacy} versus our Similarity Neighbors method. Left: empirical epsilon under DP-SGD on CIFAR-10, CIFAR-100, and CINIC-10, with our method 1.15x, 1.13x, and 1.30x higher respectively. Middle: TPR at 0.1\% FPR for undefended models on CIFAR-10 and CINIC-10, with our method 3.37x and 2.85x higher. Right: TPR at 0.1\% FPR under RelaxLoss on CIFAR-10 and CINIC-10, with our method 19.16x and 3.12x higher. In all panels, our approach yields higher leakage estimates than the baseline.}
    \label{fig:steinke_vs_sim}
\end{figure*}

\begin{figure*}[h!t!]
    \centering
\includegraphics[width=0.85\textwidth]{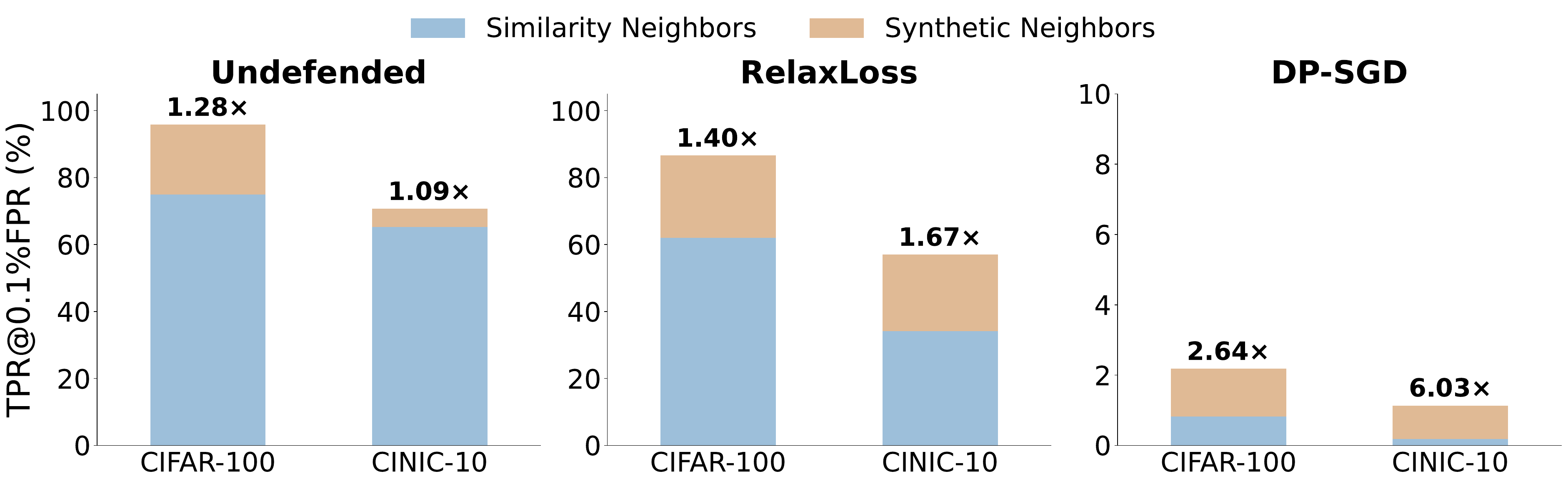}
    \caption{Settings where our Synthetic Neighbors approach significantly outperforms our Similarity Neighbors approach.}
    \Description{Three grouped bar charts comparing privacy leakage estimates from our Similarity Neighbors method versus our Synthetic Neighbors method. Left: TPR at 0.1\% FPR for undefended models on CIFAR-100 and CINIC-10, with Synthetic Neighbors 1.28x and 1.09x higher. Middle: TPR at 0.1\% FPR under RelaxLoss on CIFAR-100 and CINIC-10, with Synthetic Neighbors 1.40x and 1.67x higher. Right: TPR at 0.1\% FPR under DP-SGD on CIFAR-100 and CINIC-10, with Synthetic Neighbors 2.64x and 6.03x higher. In all panels, Synthetic Neighbors yields higher leakage estimates than Similarity Neighbors.}
    \label{fig:sim_vs_synth}
\end{figure*}

\subsection{Datasets, Defenses and Metrics}
We conduct experiments on CIFAR-10, CIFAR-100, and CINIC-10~\cite{darlow2018cinic10imagenetcifar10}, three widely used image-classification benchmarks in the privacy literature. 
To increase the distributional difference between CIFAR-10 and CINIC-10, we remove all CIFAR-10 images from CINIC-10, as their presence would otherwise introduce overlap between the two datasets and reduce the diversity of our evaluation. 
We then subsample the remaining data to 50k training and 10k test examples, matching the scale of CIFAR-10 while preserving class balance.

Prior auditing work primarily focuses on estimating the differential privacy parameter $\epsilon$ and therefore evaluates attacks exclusively against DP-SGD \cite{steinke2023privacy}. 
Our objective, however, is not to obtain the tightest privacy guarantees, but rather to assess the quality of membership scores that can be computed without auxiliary models. 
Under this perspective, privacy leakage serves as a proxy for score quality rather than an end goal in itself. 
Consequently, the evaluation need not to be restricted to $\epsilon$ estimation or to differentially private models. 
We therefore consider three privacy regimes: undefended training, RelaxLoss~\cite{chen2022relaxlossdefendingmembershipinference}, and DP-SGD, representing no protection, empirical leakage mitigation, and formal privacy guarantees, respectively.

To evaluate privacy leakage, we report TPR at 0.1\% FPR in all settings and, for DP-SGD models, the empirical $\epsilon$. Following \citet{carlini2022lira}, we adopt TPR at low FPR because low false-positive rates are critical in realistic membership inference scenarios and provide a direct measure of attack effectiveness. For DP-SGD, we additionally estimate $\epsilon$ using the procedure of \citet{steinke2023privacy}, evaluating all pairs $k_+,k_- \in {10,20,\dots,250}$, retaining the tightest estimate for each model, and reporting the average across the 64-model ensemble.

Since TPR at low FPR considers only the highest-scored members, while $\epsilon$  estimation aggregates over both members and non-members, the two metrics can diverge. Intuitively, $\epsilon$ captures average-case leakage: it reflects how well the attack discriminates overall across the score distribution. TPR at low FPR, by contrast, is a worst-case metric: it measures whether the attack can confidently identify \emph{any} members, regardless of its behavior on the rest. To see why these can disagree, consider two attacks with identical $\epsilon$ estimates: one concentrates its correct predictions at the top of the ranking (high TPR at low FPR), the other at the bottom (near-zero TPR at low FPR). This asymmetry motivates reporting both metrics jointly in DP settings.

\subsection{Similarity Neighbors}
\label{subsec:similarity_neighbors_results}
In Figure~\ref{fig:steinke_vs_sim}, we present representative settings where our Similarity Neighbors approach significantly outperforms \citet{steinke2023privacy}.  Looking at both metrics, the results follow  a consistent trend. 
Under DP-SGD , our attack achieves modest but meaningful improvements over the baseline in terms of empirical $\epsilon$ across all three datasets, suggesting that our scores are consistently stronger in the traditional one-round DP auditing setting. 
The TPR results reveal a particularly pronounced effect.
Against undefended models, our attack substantially outperforms the baseline across both CIFAR-10 and CINIC-10, indicating that our method is far more effective at identifying members when no empirical defense is active. 
The gap becomes even more dramatic under RelaxLoss, where our attack achieves almost $20 \times$ higher TPR on CIFAR-10, demonstrating that our approach is considerably more robust to this defense than the baseline.

Overall, these results highlight the importance of per-sample calibrated scores. 
They further confirm that strong calibrated signals can be effectively achieved within a one-round setting through neighbor-based approximations relying solely on test data---without the need for shadow model training or auxiliary data collection.

\subsection{Synthetic Neighbors}
\label{subsec:synthetic_neighbors_results}
Figure~\ref{fig:sim_vs_synth} reports a selection of settings in which Synthetic Neighbors outperforms Similarity Neighbors, focusing on the cases where the performance gap is most pronounced. 
Against undefended models, Synthetic Neighbors already yields a substantial advantage on CIFAR-100, reaching nearly 96\% TPR, while the gap narrows on CINIC-10. 
Under RelaxLoss, Similarity Neighbors suffer a marked degradation across both datasets, whereas Synthetic Neighbors maintain considerably higher TPR, indicating greater resilience to this form of defense. 
The contrast is most striking under DP-SGD: despite both methods being heavily suppressed by the strong privacy guarantees, Synthetic Neighbors achieve disproportionately higher TPR across both datasets, pointing to a qualitative difference in attack effectiveness.

\begin{figure*}[h!t!]
    \centering
\includegraphics[width=0.85\textwidth]{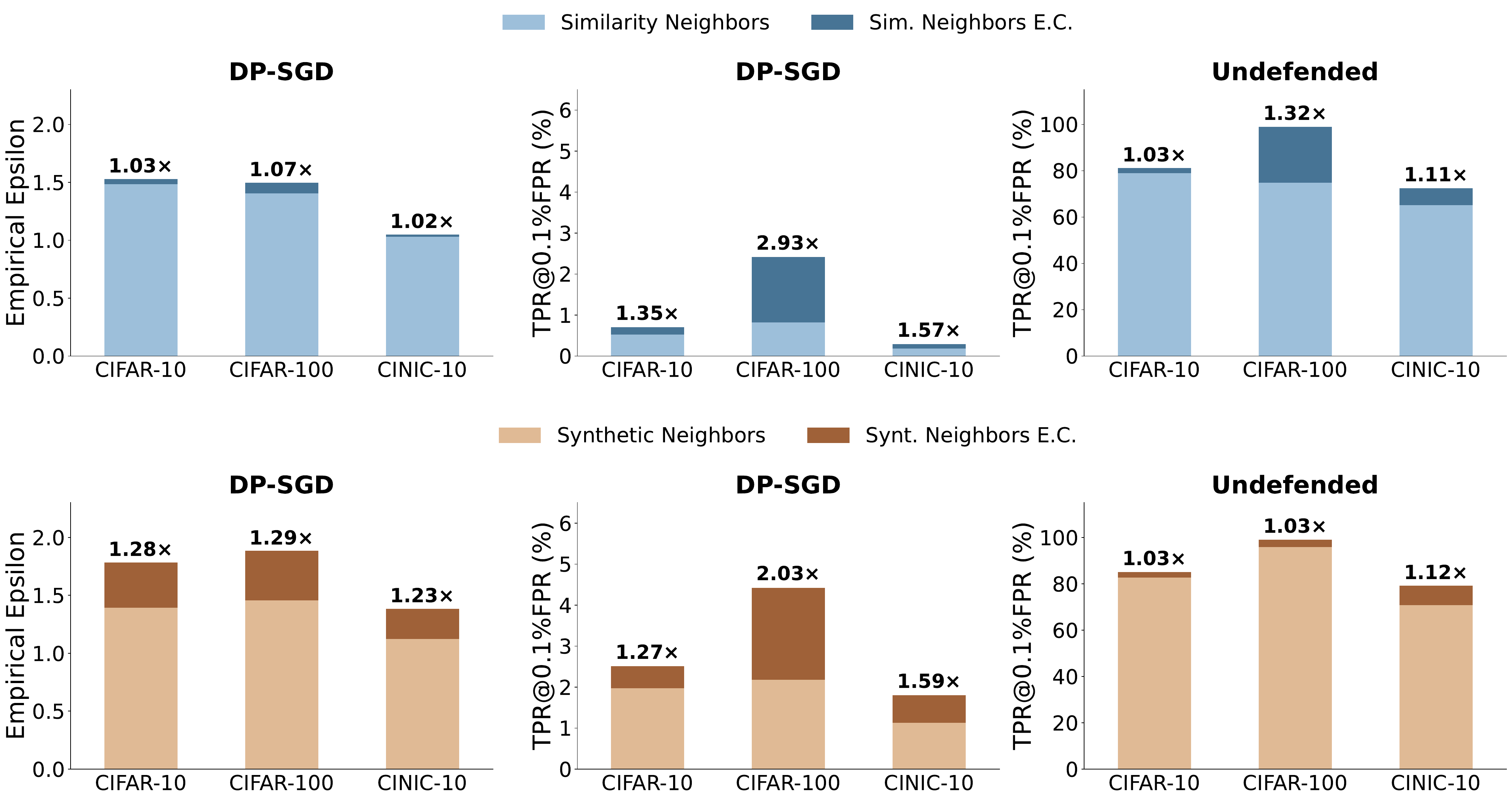}
    \caption{Settings where our Early Checkpoint (E.C.) variants significantly outperform the standard attacks.}
    \Description{Two rows of three grouped bar charts comparing standard attacks against early-checkpoint (E.C.) variants, for CIFAR-10, CIFAR-100, and CINIC-10. Top row: Similarity Neighbors versus Similarity Neighbors E.C. Left panel shows empirical epsilon under DP-SGD, with E.C. 1.03x, 1.07x, and 1.02x higher across the three datasets. Middle panel shows TPR at 0.1\% FPR under DP-SGD, with E.C. 1.35x, 2.93x, and 1.57x higher. Right panel shows TPR at 0.1\% FPR for undefended models, with E.C. 1.03x, 1.32x, and 1.11x higher. Bottom row: Synthetic Neighbors versus Synthetic Neighbors E.C., same three panel types. Left panel shows empirical epsilon under DP-SGD, with E.C. 1.28x, 1.29x, and 1.23x higher. Middle panel shows TPR at 0.1\% FPR under DP-SGD, with E.C. 1.27x, 2.03x, and 1.59x higher. Right panel shows TPR at 0.1\% FPR for an undefended model, with E.C. 1.03x, 1.03x, and 1.12x higher. In all panels, the early-checkpoint variant yields higher leakage estimates than the standard attack.}
    \label{fig:std_vs_ec}
\end{figure*}

These results highlight a consistent and defense-agnostic advantage of Synthetic Neighbors, which we attribute to the flexibility of the generative approach. 
Unlike Similarity Neighbors, whose effectiveness is inherently limited by the availability of similar samples in the test set, Synthetic Neighbors can produce informative neighbors for virtually any canary,  making the attack robust to cases where the test distribution fails to provide neighbors of adequate quality.

\subsection{Early Checkpoints}
\label{subsec:early_checkpoints_results}
The results reported in Figure~\ref{fig:std_vs_ec} consistently demonstrate that the Early Checkpoint (E.C.) variants outperform their standard counterparts across both DP-SGD and undefended settings, both metrics, and all three datasets.

For Similarity Neighbors, empirical epsilon improvements under DP-SGD are modest, with the E.C. variant providing only marginal gains across all datasets. 
The picture changes more clearly when looking at TPR at 0.1\% FPR, where the E.C. variant achieves notably stronger results, particularly on CIFAR-100. 
In the undefended setting, the improvement is most dramatic on CIFAR-100, where the standard variant reaches 74.9\% while the E.C. variant nearly saturates at 99.0\%, with more moderate but still consistent gains on CIFAR-10 and CINIC-10.

For Synthetic Neighbors, the E.C. variant delivers more substantial empirical epsilon improvements than in the similarity-based case, with consistent gains across all three datasets. 
TPR improvements under DP-SGD are consistent across all datasets, with CIFAR-100 again showing the largest relative gain. 
In the undefended setting, the E.C. variant approaches saturation on CIFAR-100 and delivers meaningful gains on CINIC-10, while CIFAR-10 shows a smaller margin due to the standard variant already performing strongly there.

Taken together, these results make a compelling case for the E.C. variants. 
The gains are not isolated to a specific attack family, dataset, or defense regime — they are systematic. 
The effect is especially pronounced in the most challenging settings,  precisely where standard attacks struggle most. 
This robustness across diverse experimental conditions strongly supports querying neighbors against early checkpoints as a broadly effective strategy, consistently yielding stronger membership inference signals regardless of the underlying neighbor generation method.

\subsection{Main Results} \label{sec:results}
Considering the results reported in the Sections~\ref{subsec:similarity_neighbors_results}, \ref{subsec:synthetic_neighbors_results} and \ref{subsec:early_checkpoints_results}, Synthetic Neighbors E.C. emerges as our strongest proposal. 
In Tables~\ref{tab:main_results_eps} and \ref{tab:main_results_tpr}, we therefore expand the direct comparison with \citet{steinke2023privacy}, already anticipated in Figure~\ref{fig:steinke_vs_synth_ec}. 
We further report results for LiRA, as upper bound on our method.

In Table~\ref{tab:main_results_eps}, we measure the privacy leakage against DP-SGD in terms of $\epsilon$, showing that our attack yields improvements over the baseline ranging from roughly 0.5 to 0.6 across the datasets, while consistently retaining more than 50\% of the LiRA performance.

In Table~\ref{tab:main_results_tpr}, we measure the TPR at 0.1\% FPR across all nine settings.
Against undefended models, we exhibit only marginal performance loss compared to LiRA, outperforming \citet{steinke2023privacy} by a factor of $\sim3.5 \times$ on both CIFAR-10 and CINIC-10.
On CIFAR-100, the higher task complexity limits the maximum achievable test accuracy, leading to a more pronounced train-test accuracy gap, which makes overfitting particularly pronounced and drives nearly all attacks to saturate near 100\% TPR.
Against DP-SGD, despite the restricted privacy leakage, our approach still outperforms the baseline by a factor of approximately $4 \times$ on CIFAR-10, $2 \times$ on CIFAR-100, and $5 \times$ on CINIC-10.
Against RelaxLoss, our method again substantially outperforms the baseline across all datasets, by a factor of ~20× on CIFAR-10 and ~5× on CINIC-10, while achieving an intermediate TPR between it and LiRA on CIFAR-100.

Overall, Synthetic Neighbors E.C. consistently and significantly closes the gap to the LiRA upper bound across all privacy settings and datasets, substantially outperforming \citet{steinke2023privacy} in every setting. 
These results highlight that more powerful and calibrated MI scores directly translate into tighter empirical lower bounds on $\epsilon$ in the context of one-round DP auditing, and, more broadly, into a more accurate estimation of the true privacy leakage of a model.

\begin{table}[t!]
\centering
\caption{Comparison between \citet{steinke2023privacy} (baseline), our Synthetic Neighbors E.C. attack and LiRA (upper bound) against DP-SGD, in terms of empirical $\epsilon$.}
\label{tab:main_results_eps}
\resizebox{\linewidth}{!}{
\begin{tabular}{ll cc|c}
    \toprule
    \multicolumn{2}{l}{\textbf{Empirical Epsilon}} & \textbf{\citet{steinke2023privacy}} & \textbf{Ours} & \textbf{LiRA} \\
    \midrule
    \multirow{3}{*}{\textbf{DP-SGD~~~~}} & CIFAR-10  & 1.29 & \textbf{1.78} & 2.99 \\
                                         & CIFAR-100 & 1.24 & \textbf{1.88} & 3.16 \\
                                         & CINIC-10  & 0.79 & \textbf{1.38} & 2.37 \\
    \bottomrule
\end{tabular}
}
\end{table}

\begin{table}[t!]
\centering
\caption{Comparison between \citet{steinke2023privacy} (baseline), our Synthetic Neighbors E.C. attack and LiRA (upper bound) across all privacy settings, in terms of TPR at 0.1\% FPR.}
\label{tab:main_results_tpr}
\resizebox{\linewidth}{!}{
\begin{tabular}{ll cc|c}
    \toprule
    \multicolumn{2}{l}{\textbf{TPR@0.1\%FPR (\%)}} & \textbf{\citet{steinke2023privacy}} & \textbf{Ours} & \textbf{LiRA} \\
    \midrule
    \multirow{3}{*}{\textbf{Undefended}} & CIFAR-10  & 23.38 & \textbf{84.98} & 97.03 \\
                                         & CIFAR-100 & \textbf{99.28} & 99.03 & 99.73 \\
                                         & CINIC-10  & 22.88 & \textbf{79.10} & 91.29 \\

    \midrule
    \multirow{3}{*}{\textbf{DP-SGD}}     & CIFAR-10  & 0.65 & \textbf{2.51} & 12.21 \\
                                         & CIFAR-100 & 2.19 & \textbf{4.43} & 13.89 \\
                                         & CINIC-10  & 0.33 & \textbf{1.80} & 4.43  \\
    \midrule

    \multirow{3}{*}{\textbf{RelaxLoss}}  & CIFAR-10  & 1.06  & \textbf{21.45} & 69.24 \\
                                         & CIFAR-100 & 72.68 & \textbf{84.64} & 98.31 \\
                                         & CINIC-10  & 10.96 & \textbf{53.73} & 78.46 \\
    \bottomrule
\end{tabular}
}
\end{table}

\section{Related Work}
\subsection{Reference-Based Membership Inference Attacks}
As anticipated in Section~\ref{subsec:loss-to-shadow}, Membership inference attacks (MIAs) have traditionally relied on shadow models—--also referred to as reference models--—as a mechanism for approximating the behavior of a target model. 
The shadow-model paradigm was introduced by \citet{shokri2017membership}, who proposed training multiple auxiliary models on datasets drawn from the same distribution as the target model and subsequently using their outputs to train an attack classifier that distinguishes members from non-members. 
This work established the canonical black-box MIA framework and demonstrated that membership information can be inferred from model predictions alone. 

Subsequent work sought to improve both the efficiency and fidelity of this paradigm. 
\citet{salem2018mlleaksmodeldataindependent} showed that effective membership inference can be achieved with significantly fewer shadow models and even without a dedicated attack classifier, suggesting that the core membership signal lies in the statistical differences between member and non-member predictions rather than in the complexity of the attack model itself.
More recently, \citet{li2022lleaksmembershipinferenceattacks} proposed $\ell$-Leaks, arguing that posterior probabilities discard valuable information and showing that shadow models trained to characterize the distribution of pre-softmax logits can substantially improve attack performance. 
These developments reflect a broader shift from heuristic attack-classifier pipelines toward statistically grounded methods that explicitly model the distributions induced by training membership. 
This evolution culminated in the Likelihood Ratio Attack (LiRA)~\cite{carlini2022lira}, which formulates membership inference as a hypothesis-testing problem. 
By explicitly modeling uncertainty and focusing on the low-false-positive-rate regime, LiRA consistently outperforms previous shadow-model-based attacks and is widely regarded as the state-of-the-art reference-model attack, making it the primary benchmark against which modern membership inference methods are evaluated.

Despite providing stronger membership signals, \emph{Online} LiRA comes at a higher computational cost than \emph{Offline} LiRA, as it relies also on \textsc{in} models---models trained on the target sample. Building on the notation introduced in Section~\ref{subsec:offline_lira}, alongside the set of \textsc{out} shadow models $\{F_1, \dots, F_n\}$, \emph{Online} LiRA trains a set of \textsc{in} shadow models $\{F'_1, \dots, F'_n\}$ to approximate the \textsc{in} score distribution as a Gaussian
\begin{equation}
  \mathcal{D}_0 \;=\; \mathcal{N}\!\left(\mu_{\mathrm{in}},\,
  \sigma_{\mathrm{in}}^2\right),
  \label{eq:in-dist}
\end{equation}
with
\begin{equation}
  \mu_{\mathrm{in}} = \frac{1}{n} \sum_{i=1}^{n} s(F'_i, x),
  \qquad
  \sigma_{\mathrm{in}}^2 = \frac{1}{n} \sum_{i=1}^{n}
  \big( s(F'_i, x) - \mu_{\mathrm{in}} \big)^2 .
  \label{eq:in-params}
\end{equation}
Membership is then assessed by the target's score likelihood ratio between $\mathcal{D}_0$ and $\mathcal{D}_1$:
\begin{equation}
  \Lambda(x) \;=\; \frac{\mathcal{N}\!\left(s(F_t, x);\mu_{\mathrm{in}},\,
  \sigma_{\mathrm{in}}^2\right)}
  {\mathcal{N}\!\left(s(F_t, x);\mu_{\mathrm{out}},\,
  \sigma_{\mathrm{out}}^2\right)}
  \label{eq:online_lira}
\end{equation}

\subsection{Reference-Free Membership Inference Attacks}
The high computational cost of shadow model training has motivated a parallel line 
of work seeking membership signals that require no reference models.
The simplest such approaches are metric-based, exploiting statistics directly 
computable from the target model. \citet{yeom2018privacy} proposed thresholding 
on the model's loss, exploiting the observation that members tend to incur lower 
loss than non-members. \citet{bertran2023scalablemembershipinferenceattacks} proposed an alternative route 
via quantile regression, training a lightweight auxiliary model to directly 
estimate the target model's score distribution without any shadow training, 
reporting competitive performance against LiRA at low false-positive rates. 

For language models specifically, \citet{carlini2021extractingtrainingdatalarge} refined 
loss-based inference by normalizing perplexity against zlib entropy, partially 
correcting for the intrinsic complexity of each sample, while token-level methods 
such as Min-K\%~\cite{shi2024detectingpretrainingdatalarge} and Min-K\%++~\cite{zhang2025minkimprovedbaselinedetecting} 
aggregate statistics over the lowest-probability tokens, grounded in the intuition 
that non-members are more likely to contain outlier tokens. Beyond general-purpose 
attacks, shadow-free methods have also been proposed for specific settings, 
including recommender systems~\cite{chi2024shadowfreemembershipinferenceattacks} and federated 
learning~\cite{deng2026efficientmembershipinferenceattacks}.

\subsection{One-Round DP Audits}
Although reference-free attacks reduce the per-attack cost, reliably estimating 
the privacy leakage---such as TPR at low FPR---still requires evaluating a large number of 
samples across multiple models. 
One-round auditing addresses this orthogonal bottleneck in DP settings by shifting the 
auditing target to a formal $\epsilon$ lower bound, constraining the access to a single trained target model. 
While averaging over multiple runs yields more representative 
estimates, a statistically valid $\epsilon$ bound can in principle be obtained from 
a single training run.

\citet{steinke2023privacy} first introduced ORA by auditing randomly included canary points into a single training run, and converting membership guesses into an $\epsilon$ lower bound via a binomial tail bound under $(\epsilon,\delta)$-DP. 
\citet{mahloujifar2024auditingfdifferentialprivacyrun} tightened this by grounding the conversion in the full $f$-DP trade-off curve and \citet{xiang2025tightprivacyauditrun} further derived the theoretically optimal membership decoder under $f$-DP via order statistics over per-canary privacy loss scores.

Despite these advances largely contributing to refine the statistical procedure that converts membership guesses to privacy lower bounds, the membership signal itself has been largely taken for granted---typically model loss or gradient projections. 
On the theoretical side, \citet{keinan2026differentialprivacyauditedrun} show that interference between canary signals is the dominant barrier to tight auditing in DP-SGD, directly motivating the design of signals with higher per-canary distinguishability.
Notably, \citet{liu2022membershipinferenceattacksexploiting} substitute the loss-based signal with a quantile regression-based membership inference attack, building on previous work from \citet{bertran2023scalablemembershipinferenceattacks}. 
Rather than relying on a global loss threshold, their method trains an auxiliary model to predict sample-specific score quantiles, yielding per-example membership decisions that better account for heterogeneity across canaries and reporting substantial gains in black-box settings. 
However, this still requires training an auxiliary model, albeit one that is independent of the target and can be kept arbitrarily small. 
In contrast, our method requires no additional model training whatsoever, bridging the gap between the statistical power of reference-based attacks and the computational efficiency demanded by one-run auditing.

\subsection{Neighbors and synthetic data in Membership Inference Attacks}
The use of neighboring or synthetic samples is not new in the MIA literature. 
Early works leveraged generative models to compensate for the lack of access to the target model's training distribution. 
For instance, \citet{yu2022blackboxlimitedquery} enrich the available data with synthetic samples, enabling the training of shadow models even when only limited information about the original training set is available. 
Beyond data augmentation, synthetic samples can also be used to probe the local behavior of a model around a target point. 
\citet{wen2023canarycoalminebettermembership} argue that a target sample alone provides limited information for membership inference. 
They therefore employ adversarial tools to optimize the generation of diverse perturbations around a canary sample, allowing for a more informative exploration of its surrounding region and improving inference performance.

This intuition has recently gained particular attention also in the context of MIAs against large language models (LLMs), where neighboring samples are used as reference points to characterize model behavior. \citet{mattern2023membershipinferenceattackslanguage} observe that reference-model-based attacks rely on the often unrealistic assumption of access to samples drawn from the training distribution. 
To overcome this limitation, they compare the model's score on the target sample with scores obtained on synthetically generated neighboring texts, eliminating the need for reference data. 
Similarly, \citet{galli2024noisyneighborsefficientmembership} exploit neighboring samples by querying \emph{noisy} variants of the target instance, obtained through stochastic perturbations in the embedding space at inference time. 
Finally, \citet{mozaffari2024semanticmembershipinferenceattack} extend this idea by generating semantically perturbed versions of the target text and training a neural classifier to distinguish members from non-members based on the model's responses across these perturbations.

\section{Limitations}
\label{sec:limitations}

As discussed in Section~\ref{sec:method}, the core difficulty underlying our approach is that the optimal distance between a canary and its neighbors depends on canary- and model-specific factors that are not known a priori, and that cannot be resolved without membership labels. We are unable to formulate a metric that reliably quantifies neighbor quality ahead of time; instead, we rely on constructions (similarity-based retrieval, diffusion strength, early checkpoints) that implicitly control this trade-off. Developing a principled, canary-specific criterion for neighbor quality remains an open problem, and we see it as the most promising direction to further close the gap to reference-based attacks.

Beyond this core limitation, each of our proposed constructions carries its own constraints. Since Similarity Neighbors are retrieved from the test set, the attack's effectiveness for a given canary is inherently limited by whether the test set contains samples that lie in an adequate similarity band around it; when no such neighbors exist, the estimated $\hat{\mathcal{D}}_1$ is a poor approximation and the attack underperforms. Synthetic Neighbors, in turn, require a generative model capable of producing suitable variations of the input, which may not be readily available or effective for all data modalities or tasks, and depend on hyperparameters whose optimal range must be explored per setting; this exploration can be more or less costly depending on how thoroughly it is performed. As noted in Section~\ref{subsec:early_checkpoints}, the Early Checkpoint variant further relaxes the black-box assumption, as it requires access to intermediate training checkpoints that are not generally available to an external attacker; it is therefore best suited to internal auditing settings, where the auditor controls training and checkpoints come at no additional cost, rather than to external attacks.

Finally, the scale of our models and experiments is limited, for two reasons. First, prior work on one-round DP auditing, including the baseline~\cite{steinke2023privacy}, evaluates on datasets and models of similar scale, so our choice keeps our results directly comparable to existing literature. Second, our evaluation protocol requires training 64 independent models across nine experimental scenarios to obtain reliable estimates of attack performance; replicating this protocol at the scale of large foundation models would demand computational resources far beyond what is feasible in this work. We leave the evaluation of our method on larger-scale models, including LLMs to future work.

\section{Conclusion}
In this work, we introduce two novel classes of black-box membership inference attacks that require no additional model training, making them suitable for one-round audits. While prior work on one-round auditing focuses on refining the procedure that estimates the privacy leakage from membership scores, we address a complementary and largely unexplored direction: the design of powerful membership scores that require no auxiliary models. We propose to approximate the \textsc{out} score distribution of Offline LiRA by querying neighbors of the canary on the target model alone, and instantiate this idea in two ways: the Similarity Neighbors attack, which retrieves neighbors from the test set using a CLIP-based similarity metric, and the Synthetic Neighbors attack, which generates neighbors via the image-to-image pipeline of Stable Diffusion. We further propose an Early Checkpoint variant of both attacks, which evaluates neighbors on intermediate snapshots of the target model to mitigate the effect of memorization on the \textsc{in} canaries. 

Our experimental evaluation across three datasets and three privacy defenses demonstrates that both proposals consistently outperform \citet{steinke2023privacy}. The Synthetic Neighbors attack and its Early Checkpoint variant emerge as the strongest attacks in virtually every setting, achieving particularly large TPR gains in the DP-SGD setting and maintaining strong performance under Undefended and RelaxLoss scenarios. The Similarity Neighbors attack, on the other hand, provides a simpler yet powerful alternative that requires no additional data generation, making it an attractive option when synthetic generation is not feasible. 

More broadly, our results suggest that meaningful approximations of state-of-the-art MIAs are achievable in one-round settings. We hope this work encourages further investigation into model-free membership inference attacks, and contributes to a more comprehensive understanding of one-round audits beyond the DP-SGD setting.
\bibliographystyle{ACM-Reference-Format}
\bibliography{references}

\appendix

\begin{table*}[h!t!]
\centering
\resizebox{\textwidth}{!}{
\begin{tabular}{l ccc ccc}
    \toprule
     & \multicolumn{3}{c}{\textbf{Empirical Epsilon}} & \multicolumn{3}{c}{\textbf{TPR@0.1\%FPR (\%)}}\\
    \cmidrule(lr){2-4} \cmidrule(lr){5-7}
    \textbf{DP-SGD} & CIFAR-10 & CIFAR-100 & CINIC-10 & CIFAR-10 & CIFAR-100 & CINIC-10 \\
    \midrule
    \textbf{\citet{steinke2023privacy}}               & 1.2867 & 1.2427 & 0.7917 & 0.6500 & 2.1875 & 0.3312 \\
    \textbf{Similarity Neighbors}      & 1.4851 & 1.4062 & 1.0305 & 0.5250 & 0.8250 & 0.1875 \\
    \textbf{Sim. Neighbors E.C.}       & 1.5294 & 1.4981 & 1.0504 & 0.7063 & 2.4188 & 0.2938 \\
    \textbf{Synthetic Neighbors}       & 1.3926 & 1.4549 & 1.1216 & 1.9688 & 2.1812 & 1.1312 \\
    \textbf{Synt. Neighbors E.C.}      & \textbf{1.7832} & \textbf{1.8837} & \textbf{1.3849} & \textbf{2.5063} & \textbf{4.4250} & \textbf{1.8000} \\
    \midrule
    \textbf{LiRA}                       & 2.9934 & 3.1560 & 2.3675 & 12.2125 & 13.8875 & 4.4250 \\
    \bottomrule
\end{tabular}
}
\caption{Experimental results against DP-SGD, across CIFAR-10, CIFAR-100 and CINIC-10. We measure both $\epsilon$ and TPR at 0.1\% FPR. Comparison between \citet{steinke2023privacy} (baseline), our Similarity and Synthetic Neighbors attacks with their Early Checkpoint (E.C.) variants and LiRA (upper bound).}
\label{tab:dpsgd_all_results}
\end{table*}

\begin{table*}[h!t!!]
\centering
\resizebox{\textwidth}{!}{
\begin{tabular}{l ccc ccc}
    \toprule
     & \multicolumn{3}{c}{\textbf{Undefended}} & \multicolumn{3}{c}{\textbf{RelaxLoss}} \\
    \cmidrule(lr){2-4} \cmidrule(lr){5-7}
    \textbf{TPR@0.1\%FPR (\%)} & CIFAR-10 & CIFAR-100 & CINIC-10 & CIFAR-10 & CIFAR-100 & CINIC-10 \\
    \midrule
    \textbf{\citet{steinke2023privacy}}               & 23.3813 & \textbf{99.2812} & 22.8813 & 1.0563 & 72.6750 & 10.9625 \\
    \textbf{Similarity Neighbors}      & 78.9062 & 74.8875 & 65.2125 & 20.2437 & 61.9687 & 34.1500 \\
    \textbf{Sim. Neighbors E.C.}       & 81.1063 & 99.0062 & 72.4125 & 20.9663 & 82.0187 & 34.4750 \\
    \textbf{Synthetic Neighbors}       & 82.6375 & 95.8625 & 70.7562 & 19.9875 & \textbf{86.6188} & \textbf{57.0500} \\
    \textbf{Synt. Neighbors E.C.}      & \textbf{84.9813} & 99.0250 & \textbf{79.1000} & \textbf{21.4500} & 84.6361 & 53.7312 \\
    \midrule
    \textbf{LiRA}                       & 97.0250 & 99.7313 & 91.2937 & 69.2375 & 98.3062 & 78.4563 \\
    \bottomrule
\end{tabular}
}
\caption{Experimental results against undefended models and RelaxLoss, across CIFAR-10, CIFAR-100 and CINIC-10. We evaluate the TPR at 0.1\% FPR. Comparison between \citet{steinke2023privacy} (baseline), our Similarity and Synthetic Neighbors attacks with their Early Checkpoint (E.C.) variants and LiRA (upper bound).}
\label{tab:undef_relloss_all_results}
\end{table*}

\section{Open Science}
The source code and scripts required to reproduce our experiments are available on GitHub at \href{https://github.com/ethz-spylab/one_round_MI}{\texttt{ethz-spylab/one\_round\_MI}}. Instructions about how to get started and the structure of the repository are contained in the \texttt{README.md} file.

In Tables \ref{tab:dpsgd_all_results} and \ref{tab:undef_relloss_all_results} we report all our experimental results.

\section{Use of Generative AI}
As stated in Section \ref{subsec:synthetic_neighbors}, we used the \href{https://huggingface.co/stable-diffusion-v1-5/stable-diffusion-v1-5}{\texttt{stable-diffusion-v1-5}} generative model to perform our attack relying on synthetic neighbors. Moreover, the writing of this paper has been assisted by LLMs.

\end{document}